\documentclass[aps,prl,floats,showpacs]{revtex4}
\usepackage{graphicx}
\begin{document}

\def\textfraction{.1}

\title{Microscopic Selection of Fluid Fingering Patterns}
\author{David A. Kessler}
\email{kessler@dave.ph.biu.ac.il}
\affiliation{Dept.  of Physics, Bar-Ilan University, Ramat-Gan, Israel}
\author{Herbert Levine}
\email{hlevine@ucsd.edu}
\affiliation{Dept. of Physics, University of California, San Diego,
9500 Gilman Drive, La Jolla, CA 92093-0319}

\date{\today}

\begin{abstract}
We study the issue of the selection of viscous fingering patterns in the limit
of small surface tension.  Through detailed simulations of anisotropic 
fingering, we demonstrate conclusively that no selection independent
of the small-scale cutoff (macroscopic selection) occurs in this system.
Rather, the small-scale cutoff completely controls the pattern, even on
short time scales, in accord with the theory of microscopic solvability.
We demonstrate that ordered patterns are dynamically
selected only for not too small surface tensions.  For extremely
small surface tensions, the system exhibits chaotic behavior and no
regular pattern is realized.
\end{abstract}


\maketitle

There has been a continuing debate regarding the
issue of the selection of viscous fingering patterns in the limit
of small surface tension. On the one hand, the 
role of surface tension in determining a unique stable steady-state finger
through the microscopic solvability mechanism \cite{kkl,langer}
has led to the belief that the details of the
small-scale restabilization control the observed patterns, notably
including the selection of a single finger filling half the
channel from a family of such fingers found initially by Saffman
and Taylor \cite{st}.
Other researchers \cite{feigenbaum,matkowsky,mineev}
have proposed various selection
criteria which do not invoke the presence of
a microscopic cut-off, and therefore may be entitled macroscopic
selection. Here, we perform detailed simulations of
anisotropic fingering and demonstrate conclusively that no macroscopic
selection occurs in this system.

\begin{figure}
\centerline{\includegraphics[width=3.in]{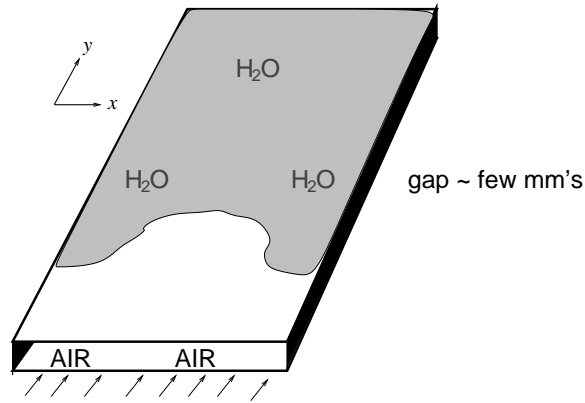}}
\caption{Schematic of the viscous fingering setup.  A Hele-Shaw cell is
filled with a viscous fluid (here water) and a less viscous fluid (here air)
is injected uniformly from one end.  The result is a finger propagating
down the channel (in the direction of increasing $y$).}
\label{setup}
\end{figure}

Viscous fingering, wherein
a low viscosity fluid invades a more viscous one in a Hele-Shaw 
channel geometry,
(see Fig.\ \ref{setup}), first described by Saffman and Taylor \cite{st}, is
one of the classic pattern forming systems and serves as the
basis for much of our intuition regarding pattern formation.
Saffman and Taylor showed both theoretically and experimentally
that this system exhibits a pattern-forming
instability, an instability that is restabilized at short length 
scales by the action
of the surface tension between the two immiscible fluids.
Detailed studies \cite{kkl,homsy,kadanoff,tanveer} of steady-state 
propagating finger solutions in this
system have revealed the following: At zero surface tension there exists
a continuous family of solutions, with continuously varying velocity and
asymmetry.  At any finite value of the surface tension, $\gamma$, 
this family is reduced to
a discrete infinity of symmetric solutions, of which only the most narrow
(and fastest) is linearly stable.  As $\gamma$ is reduced toward 
zero, all these solutions approach a width that is a fraction $\lambda=1/2$
of the channel width.  The
physical interpretation of these results has proven very controversial.
The claim has been made that it is physically unreasonable for extremely
small surface tension to ``select'' the $\lambda=1/2$ finger in times
of order 1, from a continuum of fingers whose curvatures are never 
large and should therefore all
by essentially unaffected by the surface tension.  It is been
argued that some other, macroscopic, criteria, must then be responsible for the
dynamically chosen $\lambda=1/2$ finger.  We will demonstrate in the
following that these claims are incorrect, and indeed the microscopic
regularization provided by the surface tension is dynamically
relevant and indeed controls the pattern selection.

\begin{figure}
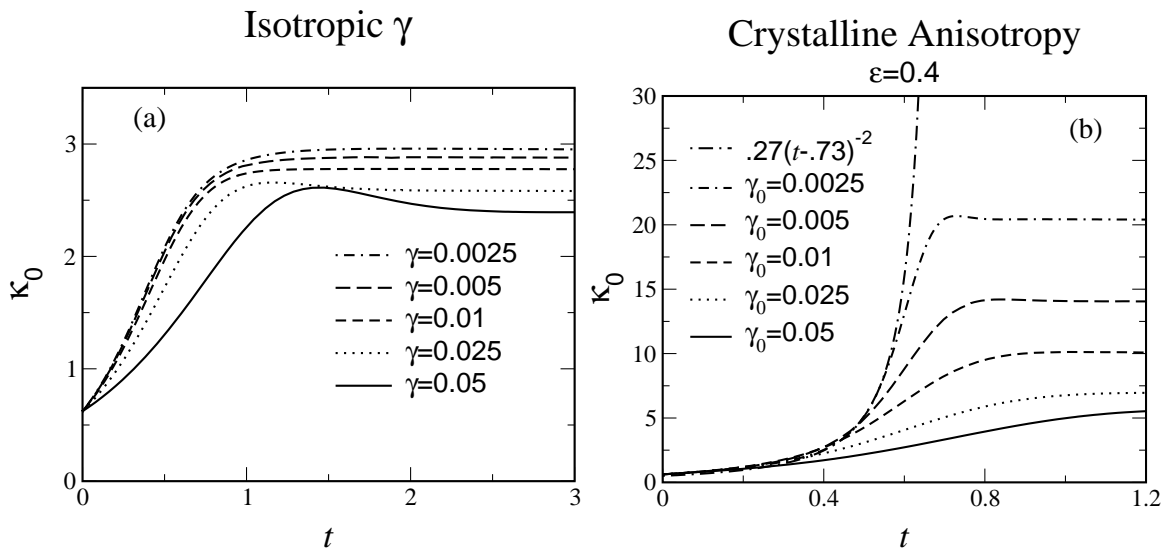

\includegraphics[width=3.in]{k_iso1_bw.eps}
\includegraphics[width=3.in]{k_anis_all_bw.eps}
\caption{(a)Tip curvature $\kappa_0$ vs. $t$ for various values of 
isotropic surface tension, $\gamma$,
starting from identical initial conditions of a small symmetric 
perturbation. (b)Same
for case of crystalline anisotropic surface tension with $\epsilon
=0.4$. The number of points on the interface, $N$, was taken between 200
and 800, depending on $\gamma$. The local relative error of the Gear's
method integrator was fixed at $10^{-6}$}
\label{sims}
\end{figure}

\begin{figure}[b]
\centerline{\includegraphics[width=3.in]{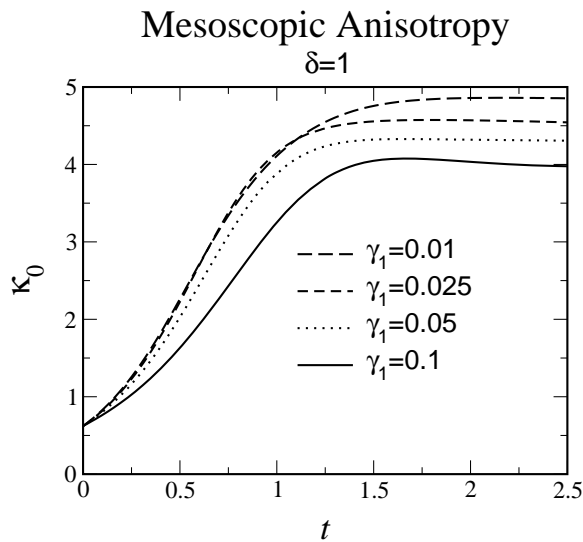}}
\caption{Simulation of $\kappa_0(t)$ using cutoff (B), 
with $\delta=1$ and varying $\gamma_1$.  The
method and initial condition are the same as in Fig.\ \ref{sims}.}
\label{meso}
\end{figure}

Part of the difficulty in unraveling the puzzle of selection is that
both proposed selection paradigms lead to the same $\lambda=1/2$ finger
at small surface tension. 
It is thus useful to consider a situation where the two mechanisms would
lead to different results.  Such is the case of viscous fingering with
a crystalline anisotropy of the surface tension. Here the boundary
condition for the pressure on the interface, $p_{\text int}$,
instead of the standard $p_{\text int}=-\gamma \kappa$,
(for surface tension $\gamma$ and 
$\kappa=-y''(x)/(1+y'(x)^2)^{3/2}$ being the curvature of the interface),
takes the form 
\begin{equation}
p_{\text int}  = -\gamma_0(1-\epsilon\cos{4\theta} ) \kappa  \quad \text{(Crystalline
anisotropy)}
\end{equation} 
where $\theta$ is the angle made by the
interface normal to the overall flow direction, $\hat y$.
This form is familiar from studies of dendritic crystal
growth \cite{kkl-pra,langer} and has already been used to model experiments with
artificially imposed anisotropy \cite{kkl-old,li,benjacob,maher}.
Study of the steady-state finger with crystalline anisotropy has
revealed \cite{kkl-pra,dorsey} that for any positive $\epsilon$, the unique stable 
finger is narrowed, with a width of order $\gamma_0^{1/2}$ for small $\gamma_0$
(a scaling identical to that of dendritic solidification of
a solid with crystalline anisotropy).
Investigation of the dynamics of such an anisotropic system should thus
reveal the competition between a macroscopic selection mechanism, if it
existed, with its favoring of the $\lambda=1/2$ finger, and the
surface tension which now favors a very narrow finger.  We do this by
solving the initial value problem (for an inviscid pushing fluid)
to see when and how the steady-state
is achieved.
In Fig.\ \ref{sims}, we present the results of this simulation, graphing the 
tip curvature, $\kappa_0$, as a function
of time for varying $\gamma$.  
The simulations were performed via the boundary integral technique
described in Refs.\ \cite{kkl-old,li}.  The interface is parametrized by 
$\theta(\alpha)$ where $0\le\alpha\le 1$ is the relative arclength, and
reflection symmetry of the finger is assumed.  The initial interface
was chosen slightly perturbed,
\begin{equation}
\label{init}
\theta(\alpha)=0.2 \sin(\pi \alpha) \ .
\end{equation}
In Fig.\ \ref{sims}a, the
results for isotropic surface tension are presented. We see that
$\kappa_0$ quickly rises, saturating at a $\gamma$ dependent value 
slightly below the value
of $\pi$ which characterizes the $\lambda=1/2$ Saffman-Taylor finger.  
(The Saffman-Taylor finger is given by the curve $
y(x) = \frac{2(1-\lambda)}{\pi} \ \ln \cos(\frac{\pi x}{2\lambda})$ ).
The selection occurs on a time scale essentially independent of $\gamma$,
leading apparent credence to the idea that the selection mechanism is
independent of $\gamma$.
In Fig.\ \ref{sims}b, however, we present the results of the same simulation, 
now performed with crystalline anisotropy, with
the anisotropy parameter fixed at $\epsilon=0.4$. If the
selection was driven by macroscopic effects,
the pattern should be
insensitive to the precise form of the surface energy for short
and intermediate times.
For such times, the data should, for short and intermediate times, 
recapitulate that of Fig.\ \ref{sims}a. Only at long times, 
times that diverge as ${\gamma_0} \rightarrow 0$,
should the narrow finger of the steady-state theory emerge. 
This is not at all what occurs. Instead, 
the selection of a narrow finger with large tip curvature 
actually is faster than for the isotropic case, with the
selection time actually decreasing with decreasing $\gamma$.
This is also consistent with
the experimental finding that imposing anisotropy creates
dendritic structures on a fast time scale \cite{benjacob}.

The generalization for dynamics of the microscopic solvability
theory of the steady state that emerges from these simulations
(see also Ref.\ \cite{prl-comment})
is that as the regularization embodied in the pressure boundary
condition is removed, the system tracks some particular
time-dependent zero-surface tension solution \cite{prl-comment}; 
but, one cannot determine which of
those solutions is selected by any macroscopic construction. The
existence of multiple trajectories emanating from arbitrarily close
initial conditions is a consequence of the Hadamard ill-posed nature of the$
\gamma=0$ problem \cite{unstable}. Fig.\ \ref{sims}b 
shows that the selected dynamics with crystalline anisotropy
is well-approximated by the curve
$\kappa_0 (t) \simeq C (t^*-t)^{-2}$ until such time as
it saturates to  $\kappa_0 (\infty ) \sim
{\gamma_0}^{-\frac{1}{2}}$. This behavior is exactly what is
expected based on the above principle. As pointed out by Howison \cite{howison}
and also by Shraiman and Bensimon \cite{sb84}, the zero-surface tension
dynamics generically exhibits finite-time singularities leading to cusped
interfaces with infinite curvature. It was shown that these cusps
can be characterized as places where the conformal map $z\equiv
x+iy =F(\xi)$ which maps the exterior of the unit circle to the
actual flow domain becomes singular via having a zero of
$\frac{dF}{d\xi}$ reach the unit circle from its interior. Assume
this happens at time $t^*$ and, without loss of generality, at
zero phase; we will also assume for simplicity that there is no
tangential component of the singularity velocity, since we are dealing
with symmetric initial conditions. (It is easy
to check that this restriction changes nothing.) Near this time, we have 
\cite{howison,sb84}
\begin{equation}
F(e^{i\theta}) \simeq K \left( e^{i\theta} - \frac{t}{t^*}
\right)^2 \ .
\end{equation}
This leads to the parametric curves
\begin{eqnarray}
x(\theta) & \simeq & - K (2-t/t^*) \theta ^2 + \mbox{constant}
\nonumber \\ 
y(\theta) & \simeq & K \left( \theta(1-t/t^*)
-\theta^3 \right) 
\end{eqnarray} 
Hence, the curve
approaches the cusp structure $|x| \simeq |y|^{\frac{2}{3}}$ via
the blow-up $\kappa_0 \simeq x''(y) \simeq
C/(t^*-t)^2$. Thus, this form of anisotropic surface tension has selected the
zero-surface tension solution with a finite-time singularity, rather
than the $\lambda=1/2$ finger.  Moreover it has done so on a time
scale independent of $\gamma$.  Clearly which pattern is
dynamically selected (even on short time scales) is controlled by the 
surface tension, despite its smallness.

\begin{figure}
\includegraphics[width=3.in]{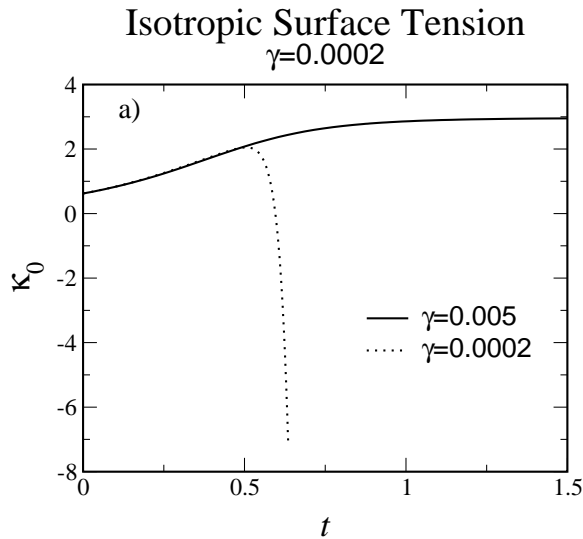}
\caption{Simulation of $\kappa_0(t)$ for isotropic surface
tension $\gamma=0.0002$ and the same initial condition
as in Fig.\ \ref{sims}.  Here $N=1000$ and the relative error $10^{-9}$.}
\label{split}
\end{figure}

We have seen that we can, by picking a suitable surface tension, cause
either the $\lambda=1/2$ or the near-cusp solution (with
tip curvature proportional to $\gamma^{-1/2}$, although
$\gamma \kappa$ at the tip is still small) to be selected.
In fact, by a suitable form of surface tension anistropy, for example,
\begin{equation}
p_{\text int} = \gamma_1 \frac{y''(x)}{(1+\delta+y'(x)^2)^{\frac{3}{2}}} \quad 
\text{(Mesoscopic anisotropy)}
\end{equation}
we can select any $\lambda (< 1/2)$
we choose.  In Fig.\ \ref{meso}, we present our results for this mesoscopic
anisotropy, 
showing that now the system chooses, in  finite time, a finger with tip
curvature which approaches a finite value significantly greater than $\pi$.
Again, there is no sign of a
macroscopically selected finger of width $\lambda$ equal to $1/2$.
To understand our
result, we note that solvability theory \cite{kkl-pra} 
predicts that the width is
determined by matching the complex-plane singularity in the
modified curvature that occurs when $y'(x) =i \sqrt{1+\delta}$
with a zero of the function $g(y') \equiv y''(y'(x))$. For the
finger, 
\begin{equation}
 g(y') = \frac{\pi (1-\lambda)}{2\lambda^2} \left( 1+
\left(\frac{\lambda y'}{1-\lambda}\right)^2 \right) 
\end{equation}
vanishes at $y'=i(1-\lambda)/\lambda$, giving the selected
values
\begin{eqnarray}
\lambda & = & \frac{1}{1+\sqrt{1+\delta}}\ ,\nonumber \\
\kappa_0 & = & \frac{\pi}{2} \sqrt{1+\delta}
(1+\sqrt{1+\delta}) \ . \end{eqnarray} 
This result is in good agreement with our simulations and rules out
any macroscopic selection.  Rather, the detailed
form of the microscopic re-stabilization indeed determines the selected
pattern.

However, by no means is this the whole story.  We saw in Fig.\ \ref{sims}a how 
reducing $\gamma$ for the usual isotropic surface tension 
appeared to yield a limit curve which asymptotes to the selected $\kappa_0=\pi$
corresponding to $\lambda=1/2$.  Let us examine what happens for an
even smaller $\gamma$ with exactly the same initial conditions.  We see in 
Fig.\ \ref{split}
that $\kappa$ starts out as before, but then turns down, going negative,
due to a splitting of the tip \cite{tip-split1,tip-split2,kkl}. 
How are we to understand this, given the
demonstration \cite{kl,bensimon,kl_physfl,tanveer_stab} 
of the linear stability of the 
selected finger for
all surface tension?  The explanation is that the size of the basin of 
attraction of the $\lambda=1/2$ steady state finger is $\gamma$ dependent,
becoming smaller (exponentially quickly) as $\gamma \to 0$ \cite{kl,bensimon}.  
\begin{figure}
\includegraphics[width=3.in]{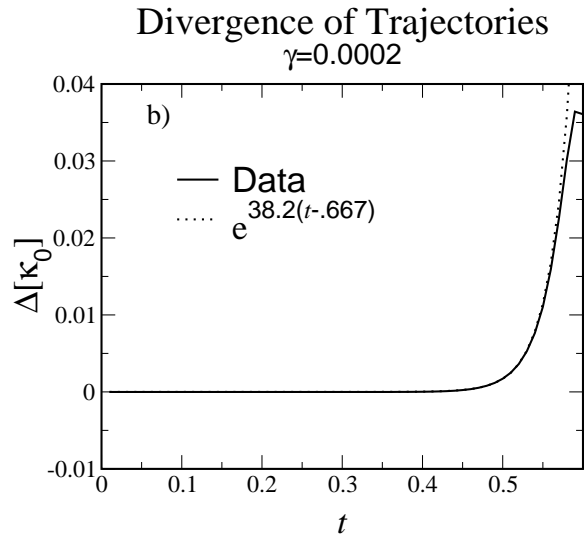}
\caption{Difference of $\kappa_0(t)$ for two runs with isotropic
surface tension $\gamma=0.0002$ and slightly different initial conditions.
Here, $N=200$ and the relative error $10^{-8}$.}
\label{lyap}
\end{figure}

Thus, the complaint
that extremely small surface tension is incapable of selecting the finger
width is perfectly correct -- at such small surface tensions, indeed, no
steady-state finger is selected.  Rather, it is likely the system is  
chaotic, generating
a random succession of tip-splittings similar to that seen in the 
Diffusion-Limited-Aggregation
model \cite{dla} of Laplacian growth, where noise is added explicitly.  
We can see this by examining
the divergence of two nearby trajectories, as shown in Fig.\ \ref{lyap}. 
Two runs
with very slightly different initial conditions were performed, the
first as in Eq. \ref{init} and the second with the
size of the initial perturbation 0.001\% larger. We find that 
initially the difference in $\kappa_0(t)$ slowly increases in a linear
fashion, but
eventually rises exponentially, increasing by over two orders of
magnitude.  Presumably, as $\gamma$ is reduced
to zero, the maximal Lyapunov exponent diverges, giving rise to the
Hadamard ill-posedness of the zero $\gamma$ problem.  Selection of the
$\lambda\approx 1/2$ finger is thus seen as a kind of intermediate
asymptotics, realized for small but not too small $\gamma$. Those searching
for a macroscopic selection principle to account for selection of $\lambda=1/2$
at extremely small surface tension are then searching for the solution
to a nonexistent dilemna.

\begin{acknowledgments}
DAK wishes to thank Prof. Lee Segel for a careful reading of the manuscript.
The work of DAK is supported by the Israel Science Foundation.  The
work of HL is supported by the National Science Foundation grant DMR-98-5735.
\end{acknowledgments}

\end{document}